\documentclass[pdflatex,sn-mathphys-num]{sn-jnl}

\usepackage{graphicx}%
\usepackage{multirow}%
\usepackage{amsmath,amssymb,amsfonts}%
\usepackage{amsthm}%
\usepackage{mathrsfs}%
\usepackage[title]{appendix}%
\usepackage{xcolor}%
\usepackage{textcomp}%
\usepackage{manyfoot}%
\usepackage{booktabs}%
\usepackage{algorithm}%
\usepackage{algorithmicx}%
\usepackage{algpseudocode}%
\usepackage{listings}%

\theoremstyle{thmstyleone}%
\theoremstyle{thmstyletwo}%

\theoremstyle{thmstylethree}%

\begin{document}

\title[Article Title]{Temporal shortcuts confound non-invasive auditory language decoding and motivate shortcut-free evaluation}


\author[1]{\fnm{Chao} \sur{Zhang}}\email{zhangchao1@pjlab.org.cn}
\equalcont{These authors contributed equally to this work.}

\author[1]{\fnm{Baoxiang} \sur{Li}}\email{libaoxiang@pjlab.org.cn}
\equalcont{These authors contributed equally to this work.}

\author[1]{\fnm{Wen} \sur{Wu}}\email{wuwen@pjlab.org.cn}

\author*[2,1,3]{\fnm{Chao} \sur{Zhang}}\email{cz277@tsinghua.edu.cn}

\affil*[1]{\orgname{Shanghai Artificial Intelligence Laboratory}, \orgaddress{\country{China}}}

\affil[2]{\orgdiv{Department of Electronic Engineering}, \orgname{Tsinghua University}, \orgaddress{\country{China}}}

\affil[3]{\orgdiv{Speech Hearing and Phonetic Sciences}, \orgname{University College London}, \orgaddress{\country{UK}}}


\abstract{
Non-invasive neural decoding of perceived speech could enable scalable brain–computer interfaces, but its evaluation often relies on temporal information unavailable during practical decoding. 
Here we identify three temporal shortcuts in electroencephalography (EEG)- and magnetoencephalography (MEG)-based word retrieval and sentence decoding.
Ground-truth word onsets reveal word duration through overlaps between adjacent input windows; variable-length sequences expose sentence word count; and absolute positional encoding reveals sentence duration. 
With the word-duration shortcut, Gaussian-noise inputs achieved a balanced top-10 accuracy of 63.2\%, closely matching the 63.7\% obtained from brain recordings. 
Gaussian-noise inputs likewise matched brain-input performance when sentence word count was available as a shortcut.
With the sentence-duration shortcut, Gaussian-noise inputs yielded a phoneme error rate of 0.244, lower than the 0.362 obtained with brain recordings.
These findings indicate that high decoding scores can be driven by temporal information embedded in input construction, independently of the recorded brain activity. 
We therefore introduce a temporal shortcut-free evaluation protocol based on frame-level word windows, fixed numbers of word embeddings and fixed-duration sentence segments, providing a fairer basis for comparing non-invasive decoders and assessing whether their performance reflects mappings between brain activity and language. Code is available at \url{https://github.com/holdmeplease/temporal-shortcut}.
}




\maketitle

\section{Introduction}\label{sec1}

Decoding language from brain activity offers a way to investigate human language processing and develop interfaces that could restore communication for people with severe speech impairments \cite{mesgarani2014phonetic,anumanchipalli2019speech}. 
Invasive speech neuroprostheses have made substantial progress, translating neural activity recorded during attempted speech into text and synthesized speech in people with paralysis \cite{willett2023high,metzger2023high}. 
However, these systems require surgical implantation of electrodes, limiting their suitability and accessibility for widespread use. 
These constraints motivate the development of non-invasive approaches to neural language decoding.

Non-invasive recordings, including electroencephalography (EEG) and magnetoencephalography (MEG), have been used to decode linguistic content during speech perception.
Contrastive approaches retrieve speech segments by matching neural recordings to speech representations, with subsequent work drawing on both acoustic and linguistic features for retrieval \cite{defossez2023decoding,boyko2024megformer,wang2026hierarchical}.
At the word level, studies have explored contextual word retrieval using convolutional–transformer architectures \cite{d2025towards} and long-context pretraining followed by fine-tuning for word retrieval \cite{jayalath2026meg}.

Research has also extended beyond retrieval to text generation and speech reconstruction.
For text generation, approaches align neural recordings with linguistic or acoustic representations to interface with pretrained language and speech models \cite{yang2026neuspeech,yang2024mad,yang2024neugpt,li2025brainecho}.
An alternative builds on word retrieval, using language-model rescoring and out-of-vocabulary infilling to assemble word predictions into sentences \cite{jayalath2025unlocking}.
For speech reconstruction, methods map neural recordings to speech waveforms, with phoneme prediction providing an additional source of supervision \cite{lee2024toward,lee2025enhancing}.
Subsequent approaches combine text and speech reconstruction or model semantic content and acoustic attributes through separate pathways before synthesizing speech \cite{jin2026neural,bao2026mindvoice}.

Despite these advances, concerns remain over whether reported decoding scores reflect the extraction of linguistic information from brain activity.
Teacher-forced evaluation, data splits with overlapping textual content, and variable-length sentence retrieval have been shown to yield high scores with limited or no advantage over noise baselines \cite{jo2025evaluating,zhang2026we,yin2025rethinking}.
These findings have prompted calls for cross-sentence split and mandatory reporting of noise-baseline performance \cite{jo2025evaluating,yin2025rethinking}.

A further concern is the use of ground-truth word onset times to align brain recordings with linguistic targets \cite{d2025towards,jayalath2025unlocking,jayalath2026meg}.
These timings are unavailable when the corresponding speech is unknown and must instead be estimated from brain activity. 
Recent MEG benchmarks and methods have investigated speech-versus-silence detection \cite{landau20252025,de2025megconformer,songyi2026mebm}, but performance on this task does not establish whether individual word onsets can be localized accurately enough for word-level decoding.

In this work, we show that timing annotations and the way model inputs are constructed introduce temporal shortcuts into auditory language decoding from EEG and MEG. 
In word retrieval, aligning input windows to ground-truth onsets makes their overlap informative about word duration, while grouping word embeddings by sentence makes the number of words available to the model. 
In sentence-generation models, absolute positional encoding can reveal sentence duration.
Across these experiments, noise inputs often yielded scores comparable to those obtained from brain recordings, and disrupting the identified cues reduced decoding performance. 
We further show that word and sentence boundaries are difficult to estimate accurately from brain activity and that errors in these estimates can degrade downstream decoding. 
On the basis of these findings, we introduce a temporal shortcut-free evaluation protocol using fixed-grid word windows, fixed numbers of word embeddings and fixed-duration signal segments.
Under this protocol, the advantage of brain recordings over matched Gaussian noise varied across datasets for word retrieval, while sentence-decoding scores showed little overall separation between the two input conditions.

\section{Results}\label{sec2}

\subsection{Word-duration shortcut}
\begin{figure}[h]
\centering
\includegraphics[width=1.0\textwidth]{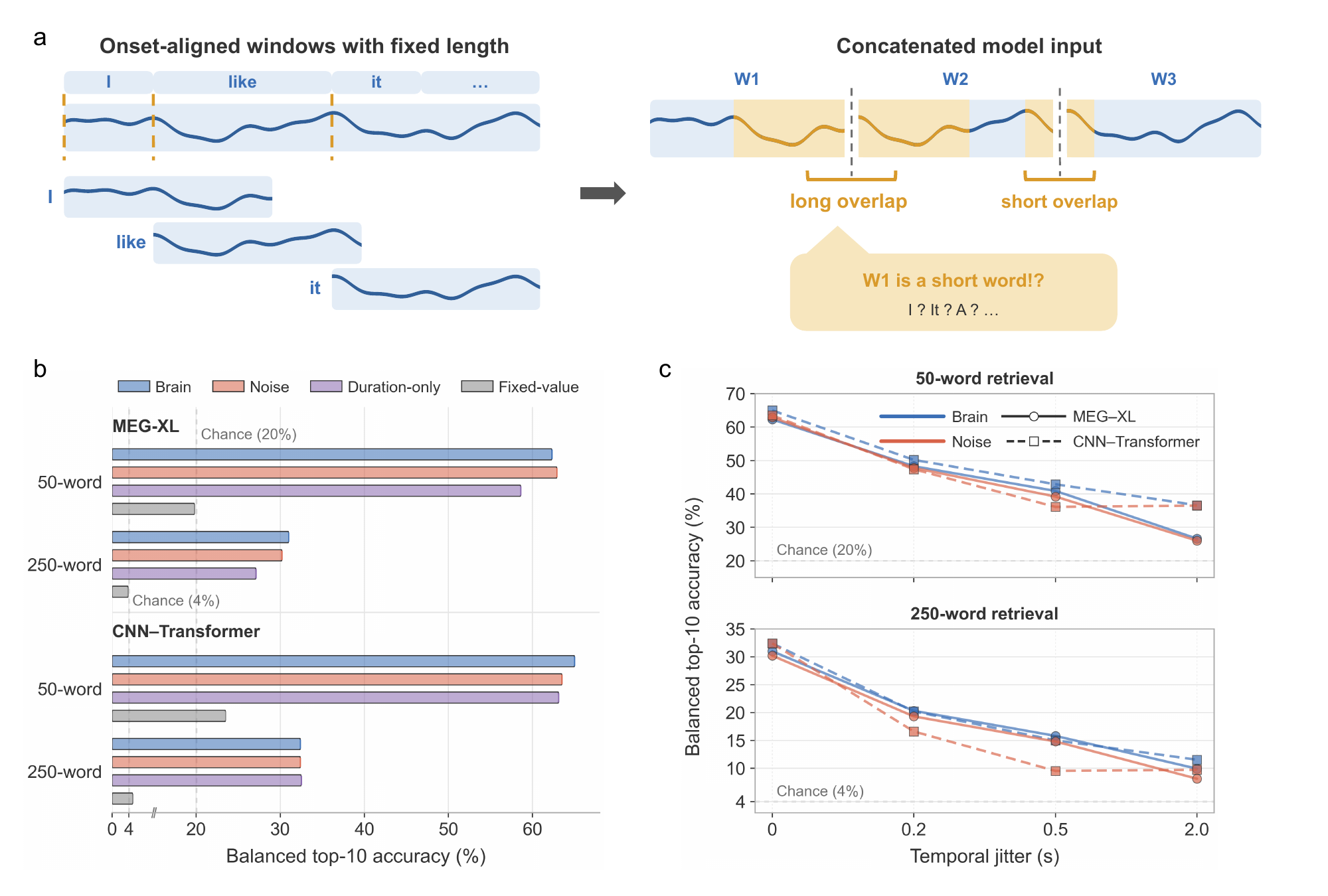}
\caption{\textbf{The word-duration shortcut in word retrieval.}
\textbf{a}, Schematic of the word-duration shortcut introduced by onset-aligned window extraction.
\textbf{b}, Balanced top-10 accuracy of MEG-XL and CNN-Transformer with brain, Gaussian-noise, duration-only and fixed-value inputs.
\textbf{c}, Retrieval accuracy as temporal jitter progressively decouples window overlap from word duration.}\label{fig1}
\end{figure}
State-of-the-art word-retrieval pipelines extract fixed-length neural windows aligned to ground-truth word onsets and concatenate windows from consecutive words before model input \cite{d2025towards,jayalath2026meg}.
Under this construction, windows associated with neighbouring words overlap in time.
The extent of this overlap is determined by the interval between successive word onsets and therefore provides the model with information closely related to word duration (Fig. 1a).

To assess the predictive value of this temporal information, we first replaced each continuous brain recording with Gaussian noise while leaving the remaining dataset structure unchanged.
Under the standard cross-sentence split, state-of-the-art models achieved balanced top-10 accuracies with Gaussian noise that were comparable to those obtained with brain recordings in both 50-word and 250-word retrieval. 
We next tested whether this non-neural performance could be explained by word duration.
Across the two evaluated models, replacing every element within each word window with the duration of the corresponding word yielded mean balanced top-10 accuracies of 60.9\% and 29.8\% in the 50-word and 250-word tasks, respectively. 
In contrast, replacing every element in every window with a single duration-independent value reduced the corresponding mean accuracies to 21.6\% and 4.4\%, close to the chance levels (Fig. 1b).
These findings indicate that word duration exposed by ground-truth onset times is sufficient to support word retrieval without brain information, providing a shortcut that does not require learning a brain-to-word mapping.

We further examined retrieval performance as precise word-duration information was progressively degraded by adding independently sampled temporal jitter to the word onsets used for window extraction. 
This perturbation shifted neighbouring windows by different amounts, progressively decoupling the extent of overlap between them from word duration.
As the jitter magnitude increased from 0 to 2\,s, balanced top-10 accuracy declined markedly for both brain and Gaussian-noise inputs in the 50-word and 250-word retrieval tasks (Fig. 1c).
These results show that current word-retrieval pipelines remain unable to decode words reliably when this information is substantially degraded.

\subsection{Sentence word-count and duration shortcut}
\begin{figure}[!h]
\centering
\includegraphics[width=1.0\textwidth]{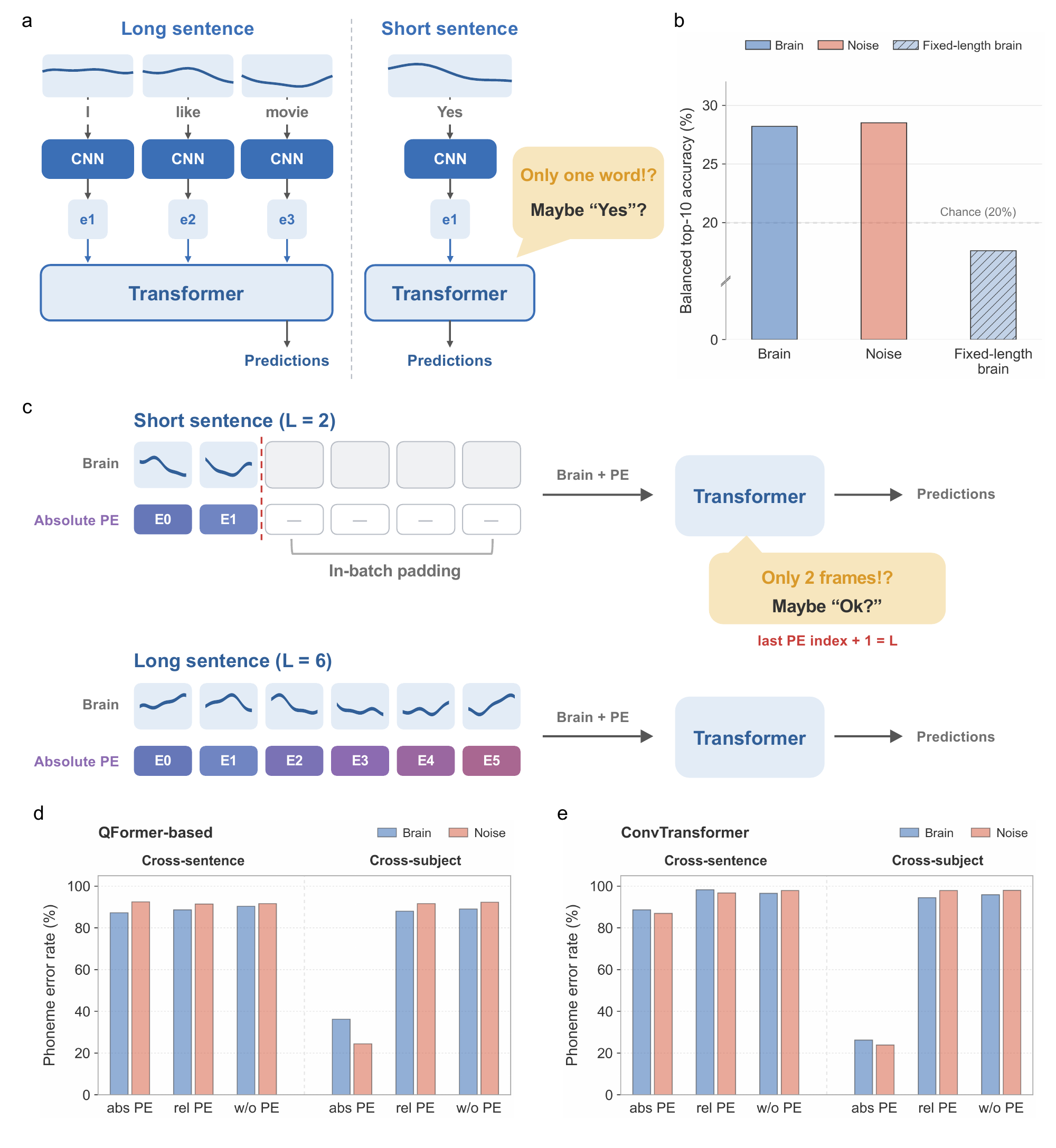}
\caption{\textbf{Sentence word-count and sentence-duration shortcuts in brain decoding.}
\textbf{a}, Schematic of the word-count shortcut in a model that represents each sentence as a variable-length sequence of word embeddings.
\textbf{b}, Balanced top-10 accuracy for brain, Gaussian-noise and fixed-length brain inputs under a CNN-Transformer architecture.
\textbf{c}, Schematic of the sentence-duration shortcut introduced when the absolute index of the final valid signal frame is exposed through positional encoding.
\textbf{d,e}, Phoneme error rate for brain and Gaussian-noise inputs with absolute, relative or no positional encoding under two different model architectures.}\label{fig2}
\end{figure}
\subsubsection{Sentence word-count shortcut}
CNN–Transformer architectures are widely used for word-level neural decoding. 
In these models, a CNN encodes each word-aligned window into a fixed-dimensional embedding, and the embeddings from consecutive words are passed to a Transformer as a sentence-level sequence.
When each input sequence corresponds to a complete sentence, the number of embeddings directly reveals the number of words in that sentence. 
Sentences of particular lengths are drawn from restricted linguistic distributions, with one-word sentences, for example, often consisting of short responses such as “yes” or “okay”.
The model can therefore use the number of input embeddings to narrow the possible word targets even when the embeddings contain little neural information. 
We refer to this dependency as the sentence word-count shortcut (Fig. 2a).

With a temporal jitter of 2\,s applied to reduce the word-duration information, we tested this shortcut by comparing natural-sentence inputs with sequences containing a fixed number of words.
Under natural sentence grouping, Gaussian-noise inputs achieved retrieval accuracy comparable to that obtained with brain recordings.
When the same word-level brain inputs were instead grouped into sequences of 50 consecutive words, balanced top-10 accuracy decreased from 28.2\% to 15.2\% (Fig. 2b).
This marked reduction indicates that models representing each sentence as a variable-length sequence of word embeddings can rely strongly on the resulting word-count shortcut.

\subsubsection{Sentence duration shortcut}
Sequence-to-sequence text and speech generation models commonly use Transformer-based models to process continuous sentence-level recordings. 
Variable sentence durations produce input sequences of different lengths, which are padded to a common length when forming a batch.
Positional encoding is essential for representing temporal order. 
However, when distinct embeddings are assigned according to position, the final valid signal frame carries its absolute index, allowing the model to infer sentence duration independently of the recorded signal content and creating a sentence-duration shortcut (Fig. 2c).

To more directly examine the influence of sentence duration, we supplemented the cross-sentence evaluation used above with a cross-subject split, in which all participants listened to the same corpus. 
A model with access to sentence duration could therefore associate the leaked duration with a previously observed sentence without relying on participant-specific brain activity.
Under the cross-sentence split, decoding performance remained limited for both brain and Gaussian-noise inputs (Fig. 2d), consistent with poor generalization of a duration-to-sentence mapping to unseen sentences.
Under the cross-subject split, performance of the QFormer-based model increased substantially when absolute positional encoding was used, with Gaussian-noise inputs achieving a phoneme error rate (PER) of 0.244, compared with 0.362 for brain recordings (Fig. 2d).
The same pattern was reproduced with a much simpler ConvTransformer model, indicating that the sentence-duration shortcut was not specific to a particular model architecture (Fig. 2e).
Thus, high cross-subject decoding scores were achievable without recorded brain activity, when duration information is available.

\subsection{Limitation in onset detection}
\begin{figure}[h]
\centering
\includegraphics[width=1.0\textwidth]{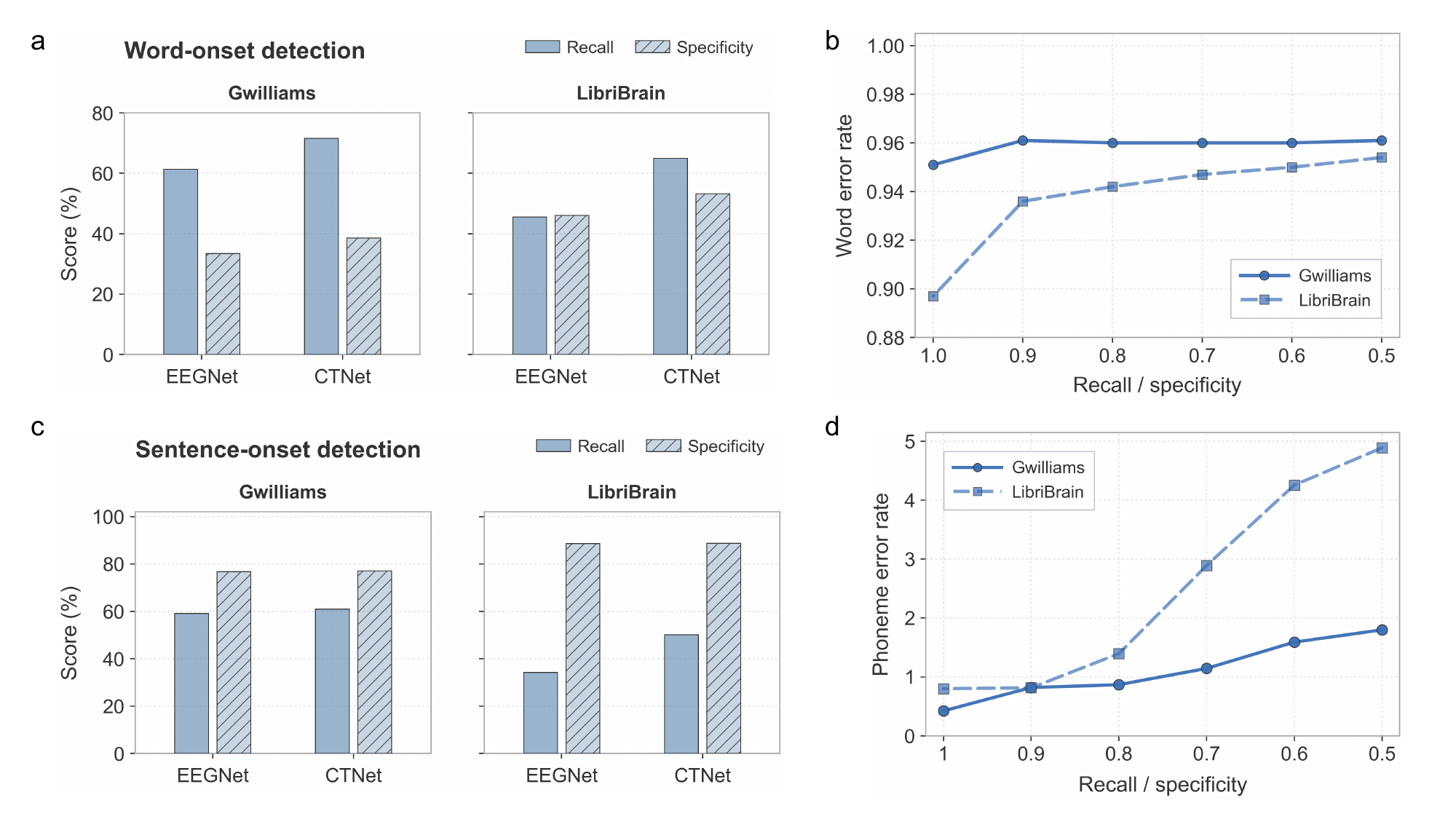}
\caption{\textbf{Onset-detection accuracy and its effect on brain decoding.}
\textbf{a}, Typical recall and specificity of word-onset detection.
\textbf{b}, Word error rate under non-ideal word onset.
\textbf{c}, Typical recall and specificity of sentence-onset detection.
\textbf{d}, Phoneme error rate under non-ideal sentence onset.}\label{fig3}
\end{figure}
Word duration and sentence duration are actually both useful information for language decoding.
We consider them as shortcuts mainly since accurately detecting word and sentence onsets from brain signals remains difficult in practical decoding systems.
As shown in Fig. 3a, word-onset detectors based on networks commonly used for EEG and MEG processing achieved a mean recall of 60.9\% and a mean specificity of 42.9\% across the two models and two datasets.
We then extracted word windows using simulated onsets under different combinations of recall and specificity and measured the resulting word error rate (WER).
Given the limited accuracy of the underlying word-retrieval model, sentence reconstruction yielded high WERs even with ground-truth word onsets, while introducing simulated onset errors further increased WER (Fig. 3b).
Across the two models and two datasets, sentence-onset detection achieved a mean recall of 51.2\% and a mean specificity of 82.8\% (Fig. 3c).
However, even at a nominal recall and specificity of 0.9, PER exceeded 0.8 on both datasets, making reliable sentence decoding difficult in practice (Fig. 3d).

\subsection{Temporal-shortcut-free evaluation protocol}
\begin{figure}[h]
\centering
\includegraphics[width=1.0\textwidth]{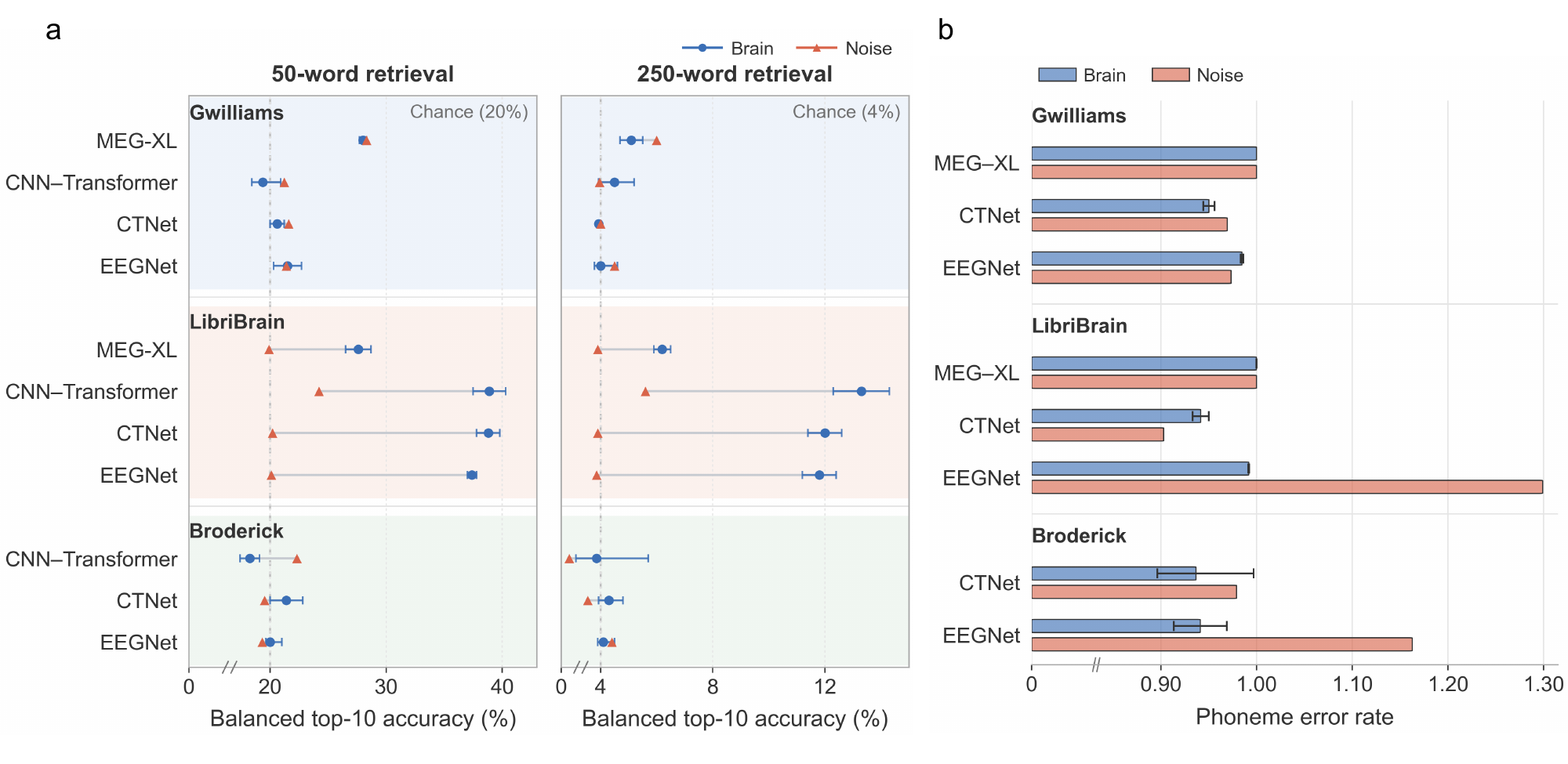}
\caption{\textbf{Temporal-shortcut-free evaluation of word retrieval and sentence decoding.}
\textbf{a}, Balanced top-10 accuracy for frame-level 50-word and 250-word retrieval with brain and Gaussian-noise inputs.
\textbf{b}, Phoneme error rate for fixed-duration sentence decoding with brain and Gaussian-noise inputs.}\label{fig4}
\end{figure}
At their core, the word-duration, sentence word-count and sentence-duration shortcuts all arise because input construction exposes information that is not reliably available during practical decoding: precise word onsets, the number of words in the current sentence and precise sentence boundaries. 
We therefore introduced a temporal-shortcut-free evaluation protocol that removes these dependencies through three design choices:
\begin{itemize}
\setlength{\itemsep}{0pt}
\setlength{\parskip}{0pt}
\item Word-level inputs are extracted using fixed frame-level windows rather than windows aligned to ground-truth word onsets.
\item Architectures combining word- and sentence-level processing receive sequences containing a fixed number of word embeddings, preventing sequence length from revealing sentence word count.
\item Sentence-level decoders receive fixed-duration signal segments rather than variable-duration inputs aligned to natural sentence boundaries.
\end{itemize}

We then evaluated existing methods using the proposed protocol, together with corresponding Gaussian-noise controls.
For word retrieval, brain and noise inputs showed little separation on Gwilliams and Broderick, while only LibriBrain retained a measurable but limited brain-noise gap (Fig. 4a).
For sentence decoding with fixed-duration inputs, brain-input PER remained high across all evaluated model--dataset combinations and did not consistently improve over the corresponding Gaussian-noise controls (Fig. 4b).
These results highlight the fundamental difficulty of decoding perceived speech from non-invasive brain recordings.

\section{Discussion}
Our analyses were limited to EEG and MEG recordings. 
Whether analogous shortcuts arise in other modalities, such as fMRI, remains unknown, and investigating their modality-specific forms will be necessary to determine how broadly our findings generalize across non-invasive brain-decoding paradigms.

This study focused primarily on identifying the mechanisms underlying temporal shortcuts and enabling fair comparison in their absence, rather than on improving brain-to-text decoding performance. 
Future work could explore decoding based on brain foundation models, which may provide more transferable neural representations from large-scale pretraining.
Repeated presentation of the same stimulus could also increase the amount of neural data available for training and improve signal quality. 
Combining these approaches with temporal-shortcut-free evaluation may help identify improvements that genuinely arise from brain activity.

\section{Conclusion}
We identified three temporal shortcuts in non-invasive auditory language decoding: word duration exposed by onset-aligned overlapping windows, sentence word count exposed by variable-length sequences of word embeddings, and sentence duration exposed through absolute positional encoding. 
In the presence of these shortcuts, Gaussian-noise inputs frequently matched or outperformed brain recordings, demonstrating that high decoding scores can arise from input construction independently of the recorded neural content. 
The limited accuracy of word- and sentence-onset detection further indicates that the temporal information required by these pipelines cannot be assumed to be available during practical decoding.
To address these limitations, we introduced a temporal-shortcut-free evaluation protocol based on frame-level word windows, fixed numbers of word embeddings and fixed-duration sentence segments. 
Re-evaluation under this protocol revealed only modest and dataset-dependent word-retrieval advantages over noise, while sentence decoding remained inaccurate. 
These findings establish a stricter basis for determining when model performance reflects brain-language mappings. 

\section{Methods}\label{sec11}

\subsection{Datasets and preprocessing}

We used the Gwilliams \cite{gwilliams2022neural} and LibriBrain \cite{ozdogan2026libribrain,mantegna2026libribrain100} MEG datasets, and the Broderick \cite{broderick2018electrophysiological} EEG dataset.
Neural recordings were band-pass filtered between 0.1 and 40\,Hz.
The filtered signals were resampled to 50\,Hz for word retrieval and 100\,Hz for sentence decoding.
Each window was baseline-corrected using the 0.5\,s of data preceding its onset, followed by robust scaling of each channel and clipping to the range $[-5,5]$.
Word- and sentence-level timing annotations were obtained from the aligned stimulus transcripts.

\subsection{Decoding tasks and data splits}

Word retrieval was evaluated using candidate vocabularies containing 50 or 250 words. 
For each input window, models ranked the candidate words, and performance was measured using balanced top-10 accuracy. 
Sentence decoding was evaluated using PER, which were computed as the Levenshtein distance normalized by the reference sequence length.


For cross-sentence evaluation, sentences were assigned to the training, validation and test sets by hashing sentence identity, targeting an 80:10:10 split. 
For cross-subject evaluation, participants were assigned to the training, validation and test sets in groups of 18, 2 and 2 for Gwilliams and 20, 4 and 8 for LibriBrain, respectively. 
Hyperparameters and checkpoints were selected using only the validation set.

\subsection{Models and training}

Word-retrieval experiments used MEG-XL \cite{jayalath2026meg}, CNN-Transformer \cite{d2025towards}, EEGNet \cite{lawhern2018eegnet} and CTNet \cite{zhao2024ctnet}.
Sentence-decoding experiments used a ConvTransformer and a variant that augments the same backbone with a QFormer module.
EEGNet and CTNet were additionally used for word- and sentence-onset detection. 
The published model architectures were retained where possible, with their input and prediction heads adapted to the corresponding datasets and tasks.

In word retrieval experiments, the MEG-XL training protocol was followed wherever applicable.
For sentence decoding, all models were trained with AdamW at a constant learning rate using connectionist temporal classification (CTC) loss, with a learning rate of $1.5\times10^{-4}$, a weight decay of 0.1 and a batch size of 256. 
In QFormer-based model, a dense token-level CLIP loss was added to the CTC loss with a weight of 1.0. 
Training was capped at 150 epochs on Gwilliams and 300 epochs on LibriBrain.

For the fixed-duration segment experiments, CTNet and EEGNet used a learning rate of $10^{-4}$ and a batch size of 256. 
MEG-XL was fine-tuned with learning rates of $10^{-5}$ for the backbone and $10^{-3}$ for the readout head, using a batch size of 64. 
Training was run for up to 50 epochs, with an early stopping patience of 10 epochs.

\subsection{Temporal-shortcut analyses}

\subsubsection{Word-duration shortcut.}

For onset-aligned word retrieval, a 3-s neural window spanning $-0.5$ to $2.5$\,s relative to each ground-truth word onset was extracted. 
Windows from consecutive words were then concatenated according to the model input format. 
For the Gaussian-noise control, each continuous EEG or MEG recording was replaced by a noise array of the same shape, generated by drawing independent samples from a standard normal distribution.
The noise was band-pass filtered using the same filter as the original recording, and the original timing annotations were retained.
For duration-only inputs, every element in a word window was replaced by a scalar representing the duration of the corresponding word. 
For fixed-value inputs, all elements were instead set to 0.5. 
Temporal jitter was applied by independently shifting each window onset by
\[
\epsilon \sim \mathcal{U}(-j,j),
\]
where $j\in\{0,0.2,0.5,2.0\}$\,s. 

\subsubsection{Sentence word-count shortcut.}

In the natural-sentence condition, the CNN encoded each word-aligned window into a fixed-dimensional embedding, and all embeddings from the same sentence were passed to the Transformer as a variable-length sequence. 
In the fixed-length condition, the same word-level inputs were reorganized into sequences containing 50 consecutive embeddings.
A temporal jitter of 2\,s was used in both conditions to reduce the contribution of the word-duration shortcut.

\subsubsection{Sentence-duration shortcut.}

Sentence-level recordings were represented as variable-length sequences and padded to the longest sequence within each batch. 
Padding positions were excluded by the attention mask.
The QFormer-based and ConvTransformer architectures were evaluated under cross-sentence and cross-subject splits with absolute, relative or no positional encoding. 
In the absolute setting, a learned embedding indexed by the absolute frame position was added to each valid frame, exposing the index of the final valid frame to the model. 
In the relative setting, each attention head had a learned bias table shared across Transformer layers. 
For a query at frame $i$ and a key at frame $j$, the bias indexed by $\operatorname{clip}(i-j,-50,50)$ was added to the attention logit. 

\subsection{Onset detection and downstream decoding}

Word- and sentence-onset detection were formulated as frame-level prediction tasks on a 50\,Hz grid. 
At each grid point, EEGNet and CTNet predicted an onset probability and a sub-grid offset that refined the estimated onset time. 
Candidate detections were thresholded at 0.5 and pruned by non-maximum suppression. 
The resulting onsets were greedily matched to reference onsets using a primary tolerance of 50\,ms.
Detection performance was reported in terms of recall and specificity.

To assess the downstream effects of word-onset errors independently of a particular detector, each reference onset was retained with probability $r$ and could generate a nearby spurious onset with probability $1-r$.
Simulated responses were smoothed before peak detection, with reference onsets used directly at $r=1$.
A 3\,s window spanning $-0.5$ to $2.5$\,s relative to each resulting onset was passed to the word-retrieval model.
The top-ranked word from each window was concatenated in onset order to form a sentence prediction, which was scored using WER.

For sentence-level decoding, onset and offset detections were sampled independently, with probability $R$ within 150\,ms of a reference boundary of the corresponding type and probability $1-S$ elsewhere. 
The sampled detections were smoothed, and the resulting onsets and offsets were greedily paired to define decoding segments.
Each segment was decoded independently using greedy CTC, and the resulting phoneme sequences were concatenated in temporal order and scored against the full reference using PER.

\subsection{Temporal-shortcut-free evaluation}

For word retrieval, the temporal-shortcut-free protocol used fixed-grid windows of 3\,s with a stride of 1.5\,s. 
Each window was labelled with the word covering its centre.
Windows whose centres were covered by no word were excluded.
Architectures that combined word- and sentence-level processing received sequences containing a fixed number of word embeddings.

For sentence-level decoding, continuous recordings were divided into non-overlapping segments of 10\,s.
The CTC target for each segment was formed by concatenating the phoneme sequences of the sentences it covered in temporal order. 
Segments spanning sentences assigned to different data splits were discarded.
Speech-free segments were retained with empty targets.

All models were evaluated with brain recordings and matched recording-level Gaussian-noise inputs. 
Unless otherwise stated, results for brain inputs were obtained across three random seeds.

\backmatter

\bibliography{bibs/neural_language,bibs/auditory_decoding,bibs/other_paradigms,bibs/evaluation,bibs/datasets,bibs/models}


\begin{thebibliography}{30}
\ifx \bisbn   \undefined \def \bisbn  #1{ISBN #1}\fi
\ifx \binits  \undefined \def \binits#1{#1}\fi
\ifx \bauthor  \undefined \def \bauthor#1{#1}\fi
\ifx \batitle  \undefined \def \batitle#1{#1}\fi
\ifx \bjtitle  \undefined \def \bjtitle#1{#1}\fi
\ifx \bvolume  \undefined \def \bvolume#1{\textbf{#1}}\fi
\ifx \byear  \undefined \def \byear#1{#1}\fi
\ifx \bissue  \undefined \def \bissue#1{#1}\fi
\ifx \bfpage  \undefined \def \bfpage#1{#1}\fi
\ifx \blpage  \undefined \def \blpage #1{#1}\fi
\ifx \burl  \undefined \def \burl#1{\textsf{#1}}\fi
\ifx \doiurl  \undefined \def \doiurl#1{\url{https://doi.org/#1}}\fi
\ifx \betal  \undefined \def \betal{\textit{et al.}}\fi
\ifx \binstitute  \undefined \def \binstitute#1{#1}\fi
\ifx \binstitutionaled  \undefined \def \binstitutionaled#1{#1}\fi
\ifx \bctitle  \undefined \def \bctitle#1{#1}\fi
\ifx \beditor  \undefined \def \beditor#1{#1}\fi
\ifx \bpublisher  \undefined \def \bpublisher#1{#1}\fi
\ifx \bbtitle  \undefined \def \bbtitle#1{#1}\fi
\ifx \bedition  \undefined \def \bedition#1{#1}\fi
\ifx \bseriesno  \undefined \def \bseriesno#1{#1}\fi
\ifx \blocation  \undefined \def \blocation#1{#1}\fi
\ifx \bsertitle  \undefined \def \bsertitle#1{#1}\fi
\ifx \bsnm \undefined \def \bsnm#1{#1}\fi
\ifx \bsuffix \undefined \def \bsuffix#1{#1}\fi
\ifx \bparticle \undefined \def \bparticle#1{#1}\fi
\ifx \barticle \undefined \def \barticle#1{#1}\fi
\bibcommenthead
\ifx \bconfdate \undefined \def \bconfdate #1{#1}\fi
\ifx \botherref \undefined \def \botherref #1{#1}\fi
\ifx \url \undefined \def \url#1{\textsf{#1}}\fi
\ifx \bchapter \undefined \def \bchapter#1{#1}\fi
\ifx \bbook \undefined \def \bbook#1{#1}\fi
\ifx \bcomment \undefined \def \bcomment#1{#1}\fi
\ifx \oauthor \undefined \def \oauthor#1{#1}\fi
\ifx \citeauthoryear \undefined \def \citeauthoryear#1{#1}\fi
\ifx \endbibitem  \undefined \def \endbibitem {}\fi
\ifx \bconflocation  \undefined \def \bconflocation#1{#1}\fi
\ifx \arxivurl  \undefined \def \arxivurl#1{\textsf{#1}}\fi
\csname PreBibitemsHook\endcsname

\bibitem[\protect\citeauthoryear{Mesgarani et~al.}{2014}]{mesgarani2014phonetic}
\begin{barticle}
\bauthor{\bsnm{Mesgarani}, \binits{N.}},
\bauthor{\bsnm{Cheung}, \binits{C.}},
\bauthor{\bsnm{Johnson}, \binits{K.}},
\bauthor{\bsnm{Chang}, \binits{E.F.}}:
\batitle{Phonetic feature encoding in human superior temporal gyrus}.
\bjtitle{Science}
\bvolume{343}(\bissue{6174}),
\bfpage{1006}--\blpage{1010}
(\byear{2014})
\end{barticle}
\endbibitem

\bibitem[\protect\citeauthoryear{Anumanchipalli et~al.}{2019}]{anumanchipalli2019speech}
\begin{barticle}
\bauthor{\bsnm{Anumanchipalli}, \binits{G.K.}},
\bauthor{\bsnm{Chartier}, \binits{J.}},
\bauthor{\bsnm{Chang}, \binits{E.F.}}:
\batitle{Speech synthesis from neural decoding of spoken sentences}.
\bjtitle{Nature}
\bvolume{568}(\bissue{7753}),
\bfpage{493}--\blpage{498}
(\byear{2019})
\end{barticle}
\endbibitem

\bibitem[\protect\citeauthoryear{Willett et~al.}{2023}]{willett2023high}
\begin{barticle}
\bauthor{\bsnm{Willett}, \binits{F.R.}},
\bauthor{\bsnm{Kunz}, \binits{E.M.}},
\bauthor{\bsnm{Fan}, \binits{C.}},
\bauthor{\bsnm{Avansino}, \binits{D.T.}},
\bauthor{\bsnm{Wilson}, \binits{G.H.}},
\bauthor{\bsnm{Choi}, \binits{E.Y.}},
\bauthor{\bsnm{Kamdar}, \binits{F.}},
\bauthor{\bsnm{Glasser}, \binits{M.F.}},
\bauthor{\bsnm{Hochberg}, \binits{L.R.}},
\bauthor{\bsnm{Druckmann}, \binits{S.}}, \betal:
\batitle{A high-performance speech neuroprosthesis}.
\bjtitle{Nature}
\bvolume{620}(\bissue{7976}),
\bfpage{1031}--\blpage{1036}
(\byear{2023})
\end{barticle}
\endbibitem

\bibitem[\protect\citeauthoryear{Metzger et~al.}{2023}]{metzger2023high}
\begin{barticle}
\bauthor{\bsnm{Metzger}, \binits{S.L.}},
\bauthor{\bsnm{Littlejohn}, \binits{K.T.}},
\bauthor{\bsnm{Silva}, \binits{A.B.}},
\bauthor{\bsnm{Moses}, \binits{D.A.}},
\bauthor{\bsnm{Seaton}, \binits{M.P.}},
\bauthor{\bsnm{Wang}, \binits{R.}},
\bauthor{\bsnm{Dougherty}, \binits{M.E.}},
\bauthor{\bsnm{Liu}, \binits{J.R.}},
\bauthor{\bsnm{Wu}, \binits{P.}},
\bauthor{\bsnm{Berger}, \binits{M.A.}}, \betal:
\batitle{A high-performance neuroprosthesis for speech decoding and avatar control}.
\bjtitle{Nature}
\bvolume{620}(\bissue{7976}),
\bfpage{1037}--\blpage{1046}
(\byear{2023})
\end{barticle}
\endbibitem

\bibitem[\protect\citeauthoryear{D{\'e}fossez et~al.}{2023}]{defossez2023decoding}
\begin{barticle}
\bauthor{\bsnm{D{\'e}fossez}, \binits{A.}},
\bauthor{\bsnm{Caucheteux}, \binits{C.}},
\bauthor{\bsnm{Rapin}, \binits{J.}},
\bauthor{\bsnm{Kabeli}, \binits{O.}},
\bauthor{\bsnm{King}, \binits{J.-R.}}:
\batitle{Decoding speech perception from non-invasive brain recordings}.
\bjtitle{Nature Machine Intelligence}
\bvolume{5}(\bissue{10}),
\bfpage{1097}--\blpage{1107}
(\byear{2023})
\end{barticle}
\endbibitem

\bibitem[\protect\citeauthoryear{Boyko et~al.}{2024}]{boyko2024megformer}
\begin{bchapter}
\bauthor{\bsnm{Boyko}, \binits{M.}},
\bauthor{\bsnm{Druzhinina}, \binits{P.}},
\bauthor{\bsnm{Kormakov}, \binits{G.}},
\bauthor{\bsnm{Beliaeva}, \binits{A.}},
\bauthor{\bsnm{Sharaev}, \binits{M.}}:
\bctitle{Megformer: enhancing speech decoding from brain activity through extended semantic representations}.
In: \bbtitle{International Conference on Medical Image Computing and Computer-Assisted Intervention},
pp. \bfpage{281}--\blpage{290}
(\byear{2024}).
\bcomment{Springer}
\end{bchapter}
\endbibitem

\bibitem[\protect\citeauthoryear{Wang et~al.}{2026}]{wang2026hierarchical}
\begin{botherref}
\oauthor{\bsnm{Wang}, \binits{B.}},
\oauthor{\bsnm{Xu}, \binits{X.}},
\oauthor{\bsnm{Xiao}, \binits{B.}},
\oauthor{\bsnm{Zheng}, \binits{L.}},
\oauthor{\bsnm{Wu}, \binits{X.}},
\oauthor{\bsnm{Cheng}, \binits{H.}},
\oauthor{\bsnm{Chen}, \binits{J.}}:
Hierarchical decoding of perceived speech from non-invasive brain recordings.
IEEE Transactions on Neural Systems and Rehabilitation Engineering
(2026)
\end{botherref}
\endbibitem

\bibitem[\protect\citeauthoryear{d’Ascoli et~al.}{2025}]{d2025towards}
\begin{barticle}
\bauthor{\bsnm{d’Ascoli}, \binits{S.}},
\bauthor{\bsnm{Bel}, \binits{C.}},
\bauthor{\bsnm{Rapin}, \binits{J.}},
\bauthor{\bsnm{Banville}, \binits{H.}},
\bauthor{\bsnm{Benchetrit}, \binits{Y.}},
\bauthor{\bsnm{Pallier}, \binits{C.}},
\bauthor{\bsnm{King}, \binits{J.-R.}}:
\batitle{Towards decoding individual words from non-invasive brain recordings}.
\bjtitle{Nature Communications}
\bvolume{16}(\bissue{1}),
\bfpage{10521}
(\byear{2025})
\end{barticle}
\endbibitem

\bibitem[\protect\citeauthoryear{Jayalath and Jones}{2026}]{jayalath2026meg}
\begin{botherref}
\oauthor{\bsnm{Jayalath}, \binits{D.}},
\oauthor{\bsnm{Jones}, \binits{O.P.}}:
Meg-xl: Data-efficient brain-to-text via long-context pre-training.
arXiv preprint arXiv:2602.02494
(2026)
\end{botherref}
\endbibitem

\bibitem[\protect\citeauthoryear{Yang et~al.}{2026}]{yang2026neuspeech}
\begin{bchapter}
\bauthor{\bsnm{Yang}, \binits{Y.}},
\bauthor{\bsnm{Duan}, \binits{Y.}},
\bauthor{\bsnm{Zhang}, \binits{Q.}},
\bauthor{\bsnm{Jo}, \binits{H.}},
\bauthor{\bsnm{Zhou}, \binits{J.}},
\bauthor{\bsnm{Lee}, \binits{W.H.}},
\bauthor{\bsnm{Xu}, \binits{R.}},
\bauthor{\bsnm{Xiong}, \binits{H.}}:
\bctitle{Neuspeech: Decode neural signal as speech}.
In: \bbtitle{ICASSP 2026-2026 IEEE International Conference on Acoustics, Speech and Signal Processing (ICASSP)},
pp. \bfpage{6636}--\blpage{6640}
(\byear{2026}).
\bcomment{IEEE}
\end{bchapter}
\endbibitem

\bibitem[\protect\citeauthoryear{Yang et~al.}{2024a}]{yang2024mad}
\begin{botherref}
\oauthor{\bsnm{Yang}, \binits{Y.}},
\oauthor{\bsnm{Jo}, \binits{H.}},
\oauthor{\bsnm{Duan}, \binits{Y.}},
\oauthor{\bsnm{Zhang}, \binits{Q.}},
\oauthor{\bsnm{Zhou}, \binits{J.}},
\oauthor{\bsnm{Hu}, \binits{X.}},
\oauthor{\bsnm{Lee}, \binits{W.H.}},
\oauthor{\bsnm{Xu}, \binits{R.}},
\oauthor{\bsnm{Xiong}, \binits{H.}}:
Mad: Multi-alignment meg-to-text decoding.
arXiv preprint arXiv:2406.01512
(2024)
\end{botherref}
\endbibitem

\bibitem[\protect\citeauthoryear{Yang et~al.}{2024b}]{yang2024neugpt}
\begin{botherref}
\oauthor{\bsnm{Yang}, \binits{Y.}},
\oauthor{\bsnm{Duan}, \binits{Y.}},
\oauthor{\bsnm{Jo}, \binits{H.}},
\oauthor{\bsnm{Zhang}, \binits{Q.}},
\oauthor{\bsnm{Xu}, \binits{R.}},
\oauthor{\bsnm{Jones}, \binits{O.P.}},
\oauthor{\bsnm{Hu}, \binits{X.}},
\oauthor{\bsnm{Lin}, \binits{C.-t.}},
\oauthor{\bsnm{Xiong}, \binits{H.}}:
Neugpt: unified multi-modal neural gpt.
arXiv preprint arXiv:2410.20916
(2024)
\end{botherref}
\endbibitem

\bibitem[\protect\citeauthoryear{Li et~al.}{2025}]{li2025brainecho}
\begin{bchapter}
\bauthor{\bsnm{Li}, \binits{J.}},
\bauthor{\bsnm{Song}, \binits{Z.}},
\bauthor{\bsnm{Wang}, \binits{J.}},
\bauthor{\bsnm{Zhang}, \binits{M.}},
\bauthor{\bsnm{Liu}, \binits{H.}},
\bauthor{\bsnm{Zhang}, \binits{M.}},
\bauthor{\bsnm{Zhang}, \binits{Z.}}:
\bctitle{Brainecho: Semantic brain signal decoding through vector-quantized spectrogram reconstruction for whisper-enhanced text generation}.
In: \bbtitle{Findings of the Association for Computational Linguistics: ACL 2025},
pp. \bfpage{2762}--\blpage{2778}
(\byear{2025})
\end{bchapter}
\endbibitem

\bibitem[\protect\citeauthoryear{Jayalath et~al.}{}]{jayalath2025unlocking}
\begin{botherref}
\oauthor{\bsnm{Jayalath}, \binits{D.}},
\oauthor{\bsnm{Landau}, \binits{G.}},
\oauthor{\bsnm{Jones}, \binits{O.P.}}:
Unlocking non-invasive brain-to-text.
In: ICML 2025 Generative AI and Biology (GenBio) Workshop
\end{botherref}
\endbibitem

\bibitem[\protect\citeauthoryear{Lee et~al.}{2024}]{lee2024toward}
\begin{botherref}
\oauthor{\bsnm{Lee}, \binits{J.}},
\oauthor{\bsnm{Kommineni}, \binits{A.}},
\oauthor{\bsnm{Feng}, \binits{T.}},
\oauthor{\bsnm{Avramidis}, \binits{K.}},
\oauthor{\bsnm{Shi}, \binits{X.}},
\oauthor{\bsnm{Kadiri}, \binits{S.}},
\oauthor{\bsnm{Narayanan}, \binits{S.}}:
Toward fully-end-to-end listened speech decoding from eeg signals.
arXiv preprint arXiv:2406.08644
(2024)
\end{botherref}
\endbibitem

\bibitem[\protect\citeauthoryear{Lee et~al.}{2025}]{lee2025enhancing}
\begin{bchapter}
\bauthor{\bsnm{Lee}, \binits{J.}},
\bauthor{\bsnm{Feng}, \binits{T.}},
\bauthor{\bsnm{Kommineni}, \binits{A.}},
\bauthor{\bsnm{Kadiri}, \binits{S.R.}},
\bauthor{\bsnm{Narayanan}, \binits{S.}}:
\bctitle{Enhancing listened speech decoding from eeg via parallel phoneme sequence prediction}.
In: \bbtitle{ICASSP 2025-2025 IEEE International Conference on Acoustics, Speech and Signal Processing (ICASSP)},
pp. \bfpage{1}--\blpage{5}
(\byear{2025}).
\bcomment{IEEE}
\end{bchapter}
\endbibitem

\bibitem[\protect\citeauthoryear{Jin et~al.}{2026}]{jin2026neural}
\begin{botherref}
\oauthor{\bsnm{Jin}, \binits{Z.}},
\oauthor{\bsnm{Li}, \binits{D.}},
\oauthor{\bsnm{Zhou}, \binits{Q.}},
\oauthor{\bsnm{Wang}, \binits{Z.}}:
Neural-vox: Neural auditory language decoding for voice and text reconstruction.
Neural Networks,
109221
(2026)
\end{botherref}
\endbibitem

\bibitem[\protect\citeauthoryear{Bao et~al.}{2026}]{bao2026mindvoice}
\begin{botherref}
\oauthor{\bsnm{Bao}, \binits{G.}},
\oauthor{\bsnm{Zeng}, \binits{T.}},
\oauthor{\bsnm{Feng}, \binits{J.}},
\oauthor{\bsnm{Xue}, \binits{X.}}:
Mindvoice: Reconstructing intelligible speech from non-invasive neural signals with pretrained priors.
arXiv preprint arXiv:2605.31173
(2026)
\end{botherref}
\endbibitem

\bibitem[\protect\citeauthoryear{Jo et~al.}{2025}]{jo2025evaluating}
\begin{barticle}
\bauthor{\bsnm{Jo}, \binits{H.}},
\bauthor{\bsnm{Yang}, \binits{Y.}},
\bauthor{\bsnm{Han}, \binits{J.}},
\bauthor{\bsnm{Duan}, \binits{Y.}},
\bauthor{\bsnm{Xiong}, \binits{H.}},
\bauthor{\bsnm{Lee}, \binits{W.H.}}:
\batitle{Evaluating eeg-to-text models through noise-based performance analysis}.
\bjtitle{Scientific Reports}
\bvolume{16}(\bissue{1}),
\bfpage{350}
(\byear{2025})
\end{barticle}
\endbibitem

\bibitem[\protect\citeauthoryear{Zhang et~al.}{2026}]{zhang2026we}
\begin{botherref}
\oauthor{\bsnm{Zhang}, \binits{X.}},
\oauthor{\bsnm{Liu}, \binits{S.}},
\oauthor{\bsnm{Lu}, \binits{R.}},
\oauthor{\bsnm{Woolgar}, \binits{A.}},
\oauthor{\bsnm{Wang}, \binits{L.}}:
What are we actually decoding? source attribution for non-invasive brain-to-language retrieval.
arXiv preprint arXiv:2605.24524
(2026)
\end{botherref}
\endbibitem

\bibitem[\protect\citeauthoryear{Yin et~al.}{2025}]{yin2025rethinking}
\begin{bchapter}
\bauthor{\bsnm{Yin}, \binits{C.}},
\bauthor{\bsnm{Yu}, \binits{Q.}},
\bauthor{\bsnm{Fang}, \binits{Z.}},
\bauthor{\bsnm{Peng}, \binits{C.}},
\bauthor{\bsnm{Li}, \binits{P.}}:
\bctitle{Rethinking cross-subject data splitting for brain-to-text decoding}.
In: \bbtitle{Proceedings of the 2025 Conference on Empirical Methods in Natural Language Processing},
pp. \bfpage{5686}--\blpage{5700}
(\byear{2025})
\end{bchapter}
\endbibitem

\bibitem[\protect\citeauthoryear{Landau et~al.}{2025}]{landau20252025}
\begin{botherref}
\oauthor{\bsnm{Landau}, \binits{G.}},
\oauthor{\bsnm{{\"O}zdogan}, \binits{M.}},
\oauthor{\bsnm{Elvers}, \binits{G.}},
\oauthor{\bsnm{Mantegna}, \binits{F.}},
\oauthor{\bsnm{Somaiya}, \binits{P.}},
\oauthor{\bsnm{Jayalath}, \binits{D.}},
\oauthor{\bsnm{Kurth}, \binits{L.}},
\oauthor{\bsnm{Kwon}, \binits{T.}},
\oauthor{\bsnm{Shillingford}, \binits{B.}},
\oauthor{\bsnm{Farquhar}, \binits{G.}}, et al.:
The 2025 pnpl competition: Speech detection and phoneme classification in the libribrain dataset.
arXiv preprint arXiv:2506.10165
(2025)
\end{botherref}
\endbibitem

\bibitem[\protect\citeauthoryear{de~Zuazo et~al.}{2025}]{de2025megconformer}
\begin{botherref}
\oauthor{\bsnm{Zuazo}, \binits{X.}},
\oauthor{\bsnm{Saratxaga}, \binits{I.}},
\oauthor{\bsnm{Navas}, \binits{E.}}:
Megconformer: Conformer-based meg decoder for robust speech and phoneme classification.
arXiv preprint arXiv:2512.01443
(2025)
\end{botherref}
\endbibitem

\bibitem[\protect\citeauthoryear{Songyi et~al.}{2026}]{songyi2026mebm}
\begin{botherref}
\oauthor{\bsnm{Songyi}, \binits{L.}},
\oauthor{\bsnm{Linze}, \binits{Z.}},
\oauthor{\bsnm{Jinghua}, \binits{L.}},
\oauthor{\bsnm{Zifeng}, \binits{Z.}}:
Mebm-speech: Multi-scale enhanced brainmagic for robust meg speech detection.
arXiv preprint arXiv:2603.02255
(2026)
\end{botherref}
\endbibitem

\bibitem[\protect\citeauthoryear{Gwilliams et~al.}{2022}]{gwilliams2022neural}
\begin{barticle}
\bauthor{\bsnm{Gwilliams}, \binits{L.}},
\bauthor{\bsnm{King}, \binits{J.-R.}},
\bauthor{\bsnm{Marantz}, \binits{A.}},
\bauthor{\bsnm{Poeppel}, \binits{D.}}:
\batitle{Neural dynamics of phoneme sequences reveal position-invariant code for content and order}.
\bjtitle{Nature communications}
\bvolume{13}(\bissue{1}),
\bfpage{6606}
(\byear{2022})
\end{barticle}
\endbibitem

\bibitem[\protect\citeauthoryear{{\"O}zdogan et~al.}{2026}]{ozdogan2026libribrain}
\begin{botherref}
\oauthor{\bsnm{{\"O}zdogan}, \binits{M.}},
\oauthor{\bsnm{Landau}, \binits{G.}},
\oauthor{\bsnm{Elvers}, \binits{G.}},
\oauthor{\bsnm{Jayalath}, \binits{D.}},
\oauthor{\bsnm{Somaiya}, \binits{P.}},
\oauthor{\bsnm{Mantegna}, \binits{F.}},
\oauthor{\bsnm{Woolrich}, \binits{M.}},
\oauthor{\bsnm{Parker~Jones}, \binits{O.}}:
Libribrain: Over 50 hours of within-subject meg to improve speech decoding methods at scale.
Advances in Neural Information Processing Systems
\textbf{38}
(2026)
\end{botherref}
\endbibitem

\bibitem[\protect\citeauthoryear{Mantegna et~al.}{2026}]{mantegna2026libribrain100}
\begin{botherref}
\oauthor{\bsnm{Mantegna}, \binits{F.}},
\oauthor{\bsnm{Jayalath}, \binits{D.}},
\oauthor{\bsnm{Elvers}, \binits{G.}},
\oauthor{\bsnm{Kim}, \binits{T.}},
\oauthor{\bsnm{Ballyk}, \binits{B.}},
\oauthor{\bsnm{Fung}, \binits{A.}},
\oauthor{\bsnm{Cho}, \binits{S.}},
\oauthor{\bsnm{Kwon}, \binits{T.}},
\oauthor{\bsnm{Kurth}, \binits{L.}},
\oauthor{\bsnm{{\"O}zdogan}, \binits{M.}}, et al.:
Libribrain100: One hundred hours of broad and deep meg data for neural speech decoding at scale.
arXiv preprint arXiv:2608.25204
(2026)
\end{botherref}
\endbibitem

\bibitem[\protect\citeauthoryear{Broderick et~al.}{2018}]{broderick2018electrophysiological}
\begin{barticle}
\bauthor{\bsnm{Broderick}, \binits{M.P.}},
\bauthor{\bsnm{Anderson}, \binits{A.J.}},
\bauthor{\bsnm{Di~Liberto}, \binits{G.M.}},
\bauthor{\bsnm{Crosse}, \binits{M.J.}},
\bauthor{\bsnm{Lalor}, \binits{E.C.}}:
\batitle{Electrophysiological correlates of semantic dissimilarity reflect the comprehension of natural, narrative speech}.
\bjtitle{Current Biology}
\bvolume{28}(\bissue{5}),
\bfpage{803}--\blpage{809}
(\byear{2018})
\end{barticle}
\endbibitem

\bibitem[\protect\citeauthoryear{Lawhern et~al.}{2018}]{lawhern2018eegnet}
\begin{barticle}
\bauthor{\bsnm{Lawhern}, \binits{V.J.}},
\bauthor{\bsnm{Solon}, \binits{A.J.}},
\bauthor{\bsnm{Waytowich}, \binits{N.R.}},
\bauthor{\bsnm{Gordon}, \binits{S.M.}},
\bauthor{\bsnm{Hung}, \binits{C.P.}},
\bauthor{\bsnm{Lance}, \binits{B.J.}}:
\batitle{{EEGNet}: a compact convolutional neural network for {EEG}-based brain--computer interfaces}.
\bjtitle{Journal of Neural Engineering}
\bvolume{15}(\bissue{5}),
\bfpage{056013}
(\byear{2018})
\end{barticle}
\endbibitem

\bibitem[\protect\citeauthoryear{Zhao et~al.}{2024}]{zhao2024ctnet}
\begin{barticle}
\bauthor{\bsnm{Zhao}, \binits{W.}},
\bauthor{\bsnm{Jiang}, \binits{X.}},
\bauthor{\bsnm{Zhang}, \binits{B.}},
\bauthor{\bsnm{Xiao}, \binits{S.}},
\bauthor{\bsnm{Weng}, \binits{S.}}:
\batitle{{CTNet}: a convolutional transformer network for {EEG}-based motor imagery classification}.
\bjtitle{Scientific Reports}
\bvolume{14}(\bissue{1}),
\bfpage{20237}
(\byear{2024})
\end{barticle}
\endbibitem

\end{thebibliography}

\end{document}